\documentclass{SCGE}
\usepackage[colorlinks,linkcolor=blue,citecolor=blue,urlcolor=blue]{hyperref}

\usepackage[utf8]{inputenc}
\usepackage{rotating}
\usepackage{footnote}
\newcolumntype{C}[1]{>{\centering\let\newline\\\arraybackslash\hspace{0pt}}m{#1}}
\usepackage[math]{blindtext}
\usepackage{times}
\usepackage{newtxtext,newtxmath}
\usepackage{graphicx}	
\usepackage{amsmath}	
\usepackage{amssymb}	
\usepackage{xspace}
\usepackage{bigints}
\usepackage{siunitx}

\usepackage{amsthm}
\usepackage{mathtools}
\usepackage[normalem]{ulem}
\usepackage{float}
\usepackage{color}
\usepackage[sort&compress,numbers]{natbib}
\usepackage{gensymb}

\usepackage{upgreek}
\usepackage{booktabs}

\usepackage{enumerate}

\let\citedash\relax
\makeatletter \providecommand{\citedash}{\hbox{-}\penalty\@m}
\makeatother
\DeclareUnicodeCharacter{2212}{-}
\begin{document}

\begin{picture}(0,0){\rm
\put(0,-20){\makebox[160truemm][l]{\bf {\sanhao\raisebox{2pt}{.}}
Article  {\sanhao\raisebox{1.5pt}{.}}}}}
\put(0,-34){\jiuwuhao {\textcolor[rgb]{0.5,0.5,0.5}{\sf 
}}}
\end{picture}

\def\bm{\boldsymbol}

\def\dl{\displaystyle}
\def\du{\end{document}}
\def\d{{\rm d}}
\def\e{{\rm e}}
\def\i{{\rm i}}

\Year{2022} %
\Month{XXXX} %
\Vol{xx} %
\No{x} %
\BeginPage{1} %
\AuthorMark{{\rm X. Shan et al.} }  
\DOI{} 
\ArtNo{000000}

\title[GWs Wave effect of microlens field]{Wave effect of gravitational waves intersected with a microlens field: \\
{\em a new algorithm and supplementary study}}

\author[1]{Xikai Shan}{}%
\author[2]{Guoliang Li}{Corresponding author (E-mail: guoliang@pmo.ac.cn)}%
\author[2,3]{Xuechun Chen}{}%
\author[2,3]{Wenwen Zheng}{}%
\author[3,4]{Wen Zhao}{}%

\address[{\rm1}]{Department of Astronomy, Beijing Normal University, Beijing 100875, China;}
\address[{\rm2}]{Purple Mountain Observatory, Chinese Academy of Sciences, Nanjing, 210023, China;}
\address[{\rm3}]{School of Astronomy and Space Science, University of Science and Technology of China, Hefei, Anhui, 230026, China;}
\address[{\rm4}]{CAS Key Laboratory for Researches in Galaxies and Cosmology, Department of Astronomy, University of Science and Technology of China, \\ Chinese Academy of Sciences, Hefei, Anhui 230026, China;}

\maketitle \vspace{-3.5mm}{\footnotesize\begin{center} Received Month date, Year; accepted Month date, Year
\end{center}}\vspace*{-5mm}

\begin{center}
\rule{16.5cm}{0.4pt}
\parbox{16.5cm}
{\begin{abstract}
The increase in gravitational wave (GW) events has allowed receiving strong lensing image pairs of GWs. However, the wave effect (diffraction and interference) due to the microlens field contaminates the parameter estimation of the image pair, which may lead to a misjudgment of strong lensing signals. To quantify the influence of the microlens field, researchers need a large sample of statistical research. Nevertheless, due to the oscillation characteristic, the Fresnel-Kirchhoff diffraction integral's computational time hinders this aspect's study. Although many algorithms are available, most cannot be well applied to the case where the microlens field is embedded in galaxy/galaxy clusters. This work proposes a faster and more accurate algorithm for studying the wave optics effect of microlenses embedded in different types of strong lensing images. Additionally, we provide a quantitative estimation criterion for the lens plane boundary for the Fresnel-Kirchhoff diffraction integral. This algorithm can significantly facilitate the study of wave optics, particularly in the case of microlens fields embedded in galaxy/galaxy clusters.
\end{abstract}}
\end{center}\vspace*{-0.6cm}

\begin{center}
\parbox{16.5cm}
{\bf\jiuhao Gravitational waves, Gravitational micro lensing, Fresnel-Kirchhoff diffraction integral}
\end{center}

\begin{center}
{\PACS{\rm 04.30.−w, 04.30.Tv, 98.62.Sb}}
\end{center}

\textwidth=178truemm \textheight=236truemm

\wuhao\vspace*{1.5mm}

\renewcommand{\baselinestretch}{1.08} \baselineskip 12.2pt\parindent=10.8pt


\section{Introduction}
Ever since advanced LIGO (aLIGO)~\citep{2015aLIGO} detected the first GW event in 2015~\citep{Abbott:2016blz}, GW as a new cosmological probe has attracted extensive attention in astrophysics. To date, aLIGO/Virgo~\citep{Acernese_2014}/KAGRA~\citep{KAGRA:2018plz} have detected $90$ GWs generated by merging compact binaries~\citep{TheLIGOScientific:2016pea, Abbott:2016blz, Abbott:2016nmj,Abbott:2017oio, TheLIGOScientific:2017qsa, Abbott:2017gyy, Abbott:2017vtc, LIGOScientific:2018mvr, abbott2021gwtc2, LIGOScientific:2021djp}. With an improvement in detector sensitivity, increasingly more GWs will be collected, allowing GWs distorted by the gravitational lenses to be identified. By comparing the wavelength of GWs with the Schwarzschild radius of the lens~\citep{2003Takahashi}, the study of the gravitational lensing of GWs can be divided into two research regimes: geometric optics, which focuses on ``short'' waves, and wave optics, which corresponds to ``long'' waves.

This study focuses on the latter case, where diffraction/interference produces a frequency-dependent distortion in the GW waveform. This phenomenon, also called the gravitational lensing wave effect, helps us determine whether GWs are gravitational lensing events by looking for such characteristic signals in a database. Many pioneering studies on the wave effect of GWs have recently been conducted in the international astronomical/physical community. Christian et al. (2018)~\cite{Christian:2018vsi} investigated the ability of aLIGO/Virgo to detect the wave effect caused by a stellar-mass lens. They found that aLIGO/Virgo could detect lenses with masses as small as $\sim30\mathrm{M}_\odot$ when the signal-to-noise ratio is $\sim30$. The Einstein Telescope and Cosmic Explorer~\citep{2014ASSL..404..333P, Evans:2016mbw} allow the detection of lenses with masses of $\sim1\mathrm{M}_\odot$. Cao et al. (2014)~\cite{PhysRevD.90.062003}, Lai et al. (2018)~\cite{Lai:2018rto}, and Jung \& Shin (2019)~\cite{PhysRevLett.122.041103} found that intermediate-mass black holes ($100\sim10^5\mathrm{M}_\odot$) could be detected through the wave effect of GWs for the aLIGO/Virgo detector network. Hannuksela et al. (2019)~\cite{2019Hannuksela} and Abbott et al. (2021c)~\cite{LIGOScientific:2021izm} searched for the wave optics events of GWs in the O1, O2, and O3a databases by approximating a point mass lens model. Nevertheless, they found no compelling evidence in the observed GWs.

In addition to the case of an isolated lens, some studies focused on microlens embedded in galaxy/galaxy clusters. Cheung et al. (2021)~\cite{Cheung:2020okf} and Yeung et al. (2021)~\cite{Yeung:2021roe} studied a point mass embedded in Type I and Type II macroimages. They found that the existence of galaxies dramatically changes the lensing phenomenon. Therefore, approximating them as an isolated point mass lens is unreasonable. Meena \& Bagla (2020)~\cite{Meena:2019ate} analyzed the influence of a point mass lens on parameter estimation for different strong lensing images and found that it can lead to detectable differences in the signal. Additionally, some works focused on the microlens field embedded in an external gravitational potential~\citep[e.g.,][]{2019Diego,2021Anuj} and showed a more distinctive appearance. These features, different from an isolated lens, indicate that studying the wave effect for microlenses embedded in galaxy/galaxy clusters is necessary. A challenging problem in this domain is calculating the Fresnel-Kirchhoff diffraction integral. Although some integration algorithms have been developed, they mainly focus on isolated lenses~\citep[e.g.,][]{press1992numerical,levin1982procedures,filon1930iii,xiang2007efficient,iserles2006computation,guo2020convergence}, etc. The specific objective of this study is to propose a faster, more accurate, and complete algorithm for all the cases of microlens/microlens fields embedded in a strong lensing image. In detail, they are Types I-III images. Additionally, the setting of the microlens field (such as a stellar field) boundary is also a significant problem that will affect the convergence of the Fresnel-Kirchhoff diffraction integral. For this aspect, we provide a quantitative estimation criterion for the first time.

This paper is organized as follows. Section~\ref{sec:Methodology} describes the basic theory of the lensing wave effect and introduces our algorithm. Section~\ref{sec:ResOnePoint} tests our algorithm for a point mass lens embedded in external convergence and compares it with the analytic results. Section~\ref{sec:ResMulMic} presents the results for microlens fields embedded in three types of macroimages. Finally, summaries and discussions are presented in Section~\ref{sec:SumDis}.

\section{Methodology}
\label{sec:Methodology}
In order to calculate the Fresnel-Kirchhoff diffraction integral faster and more accurately, we propose a new method based on Ulmer \& Goodman (1995)~\cite{UG95} and Diego et al. (2019)~\cite{2019Diego}.
This method focuses on calculating the wave effect of microlenses embedded in Type I, Type II, and Type III macro-image environments.
In section~\ref{subsec:DiffInter}, we briefly review the gravitational lensing wave effect theory.
In section~\ref{subsec:ParaSet}, we discuss the estimation criteria of the lens plane boundary.
In section~\ref{subsec:TypeI}$ \sim$ \ref{subsec:TypeIII}, we present the algorithms for different macro-image types.

\subsection{Fresnel-Kirchhoff diffraction integral}
\label{subsec:DiffInter}
In the case of scalar wave approximation, the lensing wave effect of GW can be quantified by the Fresnel-Kirchhoff diffraction integral~\citep{1992grlebookS,10.1143/PTPS.133.137,2003Takahashi}
\begin{equation}
\label{eq:DiffInter}
F(\omega, \boldsymbol{y})=\frac{2 G \mathrm{M}_{L}\left(1+z_{L}\right) \omega}{\pi c^{3} i} \int_{-\infty}^{\infty} d^{2} x \exp \left[i \omega t(\boldsymbol{x}, \boldsymbol{y})\right] ,
\end{equation}
where $F(\omega, \boldsymbol{y})$ is the magnification factor, $\omega$ and $\boldsymbol{y}$ are the circle frequency of the GW and its position in the source plane 
in the unit of the Einstein radius. 
$ \mathrm{M}_{L}$ is the lens mass, $z_L$ is the lens redshift, $\boldsymbol{x}$ is the coordinate at the lens plane, and
\begin{equation}
t(\boldsymbol{x}, \boldsymbol{y})=\frac{4 G \mathrm{M}_{L}\left(1+z_{L}\right)}{c^{3}}\left[\frac{1}{2}|\boldsymbol{x}-\boldsymbol{y}|^{2}-\psi(\boldsymbol{x})+\phi_{\mathrm{m}}(y)\right] 
\end{equation}
is the time delay function.
$\psi(\boldsymbol{x})$ is the dimensionless deflection potential, and $\phi_{\mathrm{m}}(y)$ is used to correct the zero point of the time delay.

To simplify the integral formula, we redefine the coefficient before the integrand of Eq.~(\ref{eq:DiffInter}) as
\begin{equation}
C_{\omega}=\frac{2 G \mathrm{M}_{L}\left(1+z_{L}\right) \omega}{\pi c^{3} i} .
\end{equation}
Due to the oscillation characteristics of the integrand, the traditional Gaussian numerical integration method is powerless. 
In order to solve this problem, Ulmer \& Goodman (1995)~\cite{UG95} (hereafter referred to as UG95) proposed a new integration algorithm.
In detail, one can obtain the time domain magnification factor $\tilde{F}\left(t, \boldsymbol{y}\right)$ by Fourier transforming $\frac{F(\omega, \boldsymbol{y})}{C_{\omega}}$
\begin{equation}
\label{eq:DiffTime}
\tilde{F}\left(t, \boldsymbol{y}\right) \equiv \frac{1}{2 \pi} \int_{-\infty}^{\infty} \mathrm{d} \omega \exp \left(-i \omega t\right) \frac{F(\omega, \boldsymbol{y})}{C_{\omega}}. 
\end{equation}
After inserting Eq.~(\ref{eq:DiffInter}) into Eq.~(\ref{eq:DiffTime}), one can find that the time domain magnification factor is proportional to the time delay probability density function (PDF)
\begin{equation}
\label{eq:TimeDomainMag}
\tilde{F}\left(t, \boldsymbol{y}\right) =\int_{-\infty}^{\infty} \mathrm{d}^{2} x \delta\left[t(\boldsymbol{x}, \boldsymbol{y})-t\right]=\frac{|\mathrm{d} S|}{\mathrm{d} t },
\end{equation}
where $\mathrm{d}S$ is the lens plane area, whose time delay is between $\left[t, t + \mathrm{d}t\right]$.

Then one only needs to inverse Fourier transform $\tilde{F}\left(t, \boldsymbol{y}\right)$ to get $F(\omega, \boldsymbol{y})$.
In the following text, we can find that there is no integral oscillation problem in this algorithm.
However, the inverse Fourier transformation needs to treat an infinite time delay series. 
We call it the infinite time Fourier transforming problem.
In previous works, scholars did not pay much attention to this aspect and merely used the apodization method to make a truncation.
\begin{equation}
\label{eq:apodization}
\tilde{F}_\text{apo}(t,\boldsymbol{y}) = \tilde{F}(t,\boldsymbol{y}) \times W(t),
\end{equation}
where $W(t)$ is the window function.
One can achieve different goals by choosing different window functions.
However, in the frequency domain, this operation will introduce an inevitable spectrum leakage problem due to the numerical result is the convolution of the theoretical result and the window function.
\begin{equation}
\label{eq:convolution}
F_\mathrm{apo}(\omega, \boldsymbol{y})=F(\omega,\boldsymbol{y})*W(\omega),
\end{equation}
where ``$*$'' stands for the convolution operator and $W(\omega)$ is the IFFT result of $W(t)$.
Therefore the apodization method may introduce uncontrollable errors, lose efficiency, and cause some extensive sample studies to be abandoned due to the limitation of accuracy and speed.
To address this issue, we will propose a new algorithm.
In addition, we will also give a quantitative lens plane boundary estimation criterion to balance the accuracy and speed when calculating the time domain magnification factor $\tilde{F}(t,\boldsymbol{y})$.

\subsection{Lens plane boundary estimation}
\label{subsec:ParaSet}
Due to convergence ($\kappa$) and shear ($\gamma$), the time delay of microlenses embedded in the lens galaxy/galaxy cluster is~\citep{Wambsganss1990, 1992grlebookS, 2021xuechunchen}
\begin{equation}
\begin{split}
\label{eq:TimeDelay}
t(\boldsymbol{x},\boldsymbol{x}^{i},\boldsymbol{y}=0)&=\underbrace{\frac{k}{2}\left((1-\kappa+\gamma) x_{1}^{2}+(1-\kappa-\gamma) x_{2}^{2}\right)}_{t_\text{s}(\kappa,\gamma,\boldsymbol{x})} -  \underbrace{\left[\frac{k}{2}\sum_{i}^{N} \ln \left(\boldsymbol{x}^{i}-\boldsymbol{x}\right)^{2} + k\phi_{-}(\boldsymbol{x})\right]}_{t_\text{m}(\boldsymbol{x},\boldsymbol{x}^{i})} = k\left(\tau_\text{s}(\kappa,\gamma,\boldsymbol{x})- \tau_\text{m}(\boldsymbol{x},\boldsymbol{x}^{i})\right)
\end{split}
\end{equation}
where $k=4 G \text{M}_\text{micro}(1+z_L)/c^3$ and $\boldsymbol{x^{i}}$ is coordinate of the $i$th microlens.
It is worth noting that we have set the macro image point as the coordinate origin here, so $y = 0$ (unless otherwise specified, we use $y = 0$ in the following text).
$\phi_{-}(\boldsymbol{x})$ is the contribution from a negative mass sheet which is used to
keep the total convergence $\kappa$ unchanged when adding microlenses~\citep{Wambsganss1990, 2021xuechunchen, zheng2022}.

\begin{figure}
	\centering 
	\includegraphics[width=0.73\textwidth]{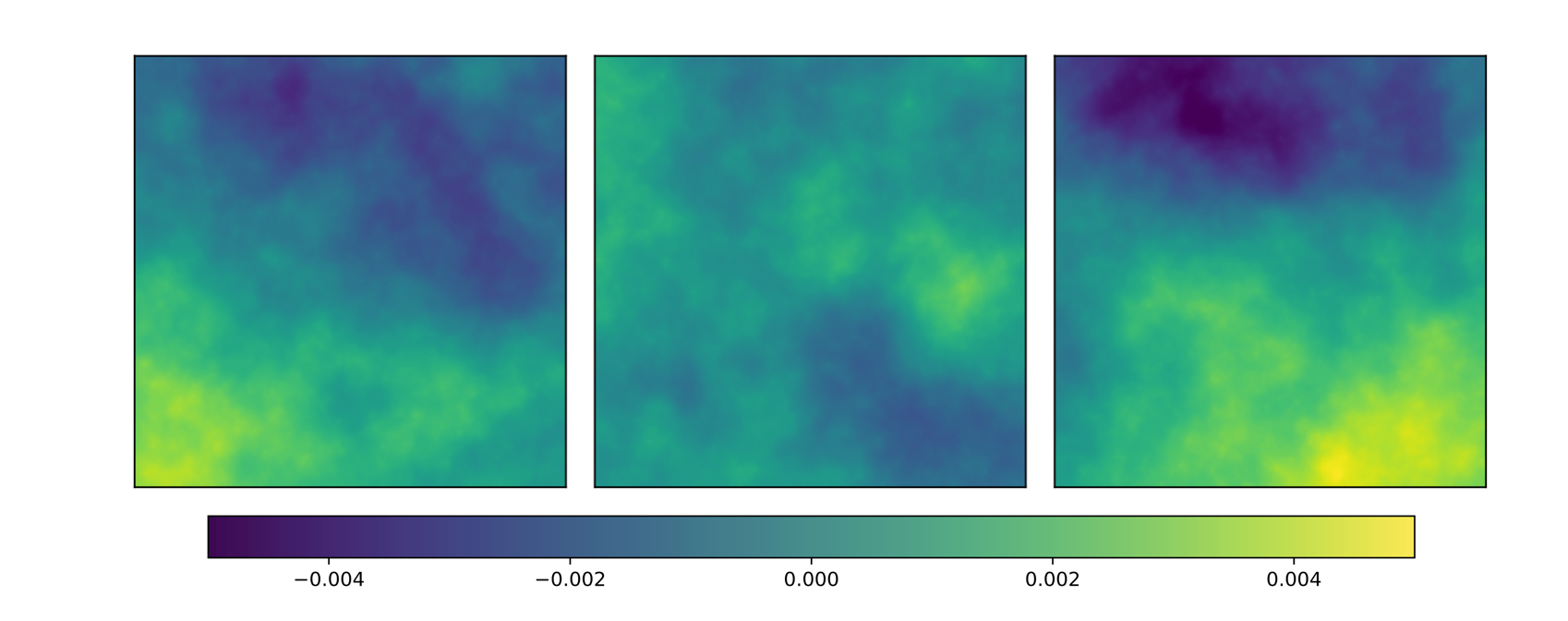}
	\caption{This figure shows three sky maps using the ``test $2$'' configuration listed in Table~\ref{ta:FluaTest}.
	The color bar represents the time delay values in seconds.}
	\label{fig:map}
\end{figure}

The term $t_\text{m}(\boldsymbol{x},\boldsymbol{x}^{i})$, which indicates the microlens time delay, is the summation of the last two terms. 
As shown in Fig.~\ref{fig:map}, this term contributes an almost uniform perturbation field to $t_\text{s}(\kappa,\gamma,\boldsymbol{x})$ which denotes the time delay from smooth potential and geometry.
So the idea here is that if we can estimate the standard deviation, $\sigma_t$ of $t_\text{m}(\boldsymbol{x},\boldsymbol{x}^{i})$ in a simulated microlens field, then we can set a smooth time delay contour $\boldsymbol{x_c}$ on which $|t_\text{s}(\kappa,\gamma,\boldsymbol{x_c})|=f\sigma_t$, $f>>1$. Beyond this contour, the time delay can be well approximated with $t_\text{s}$.
We will estimate $\sigma_t$ and discuss the corresponding lens plane boundary in the following text.

In the regime of microlenses, it is convenient to use a circular microlens field~\citep{kayser1986astrophysical,Wambsganss1990} in the polar coordinate system. Since the perturbation, $t_\text{m}$ is almost uniform, we focus on the boundary here simply.
The time delay caused by the negative mass sheet at this circular microlens field boundary $R$ is
\begin{equation}
\begin{split}
\label{eq:minusms}
\phi_{-}(R)=-\frac{\kappa_{*}}{2\pi} \int_{0}^{2\pi} \int_{0}^{R} & \ln \left[\left(R-r \cos(\theta)\right)^{2} +\left(r\sin(\theta)\right)^{2} \right] r\mathrm{d} r \mathrm{~d} \theta = - \kappa_{*} R^2 \ln{R} ,
\end{split}
\end{equation} 
where $\kappa_*$ is the convergence contributed by microlenses.
Due to the symmetry of $\phi_{-}(\boldsymbol{x})$, we use $\phi_{-}(R)$ to represent $\phi_{-}(\boldsymbol{R})$.

For only one microlens, we can replace $\kappa_{*}$ with $\frac{1}{R^2}$ to obtain $\phi_{1-}(R) = -\ln R$.
Hence, the dimensionless micro time delay induced by one microlens at the boundary $R$ is 
\begin{equation}
\tau_\text{m}(R, \tilde{r}, \tilde{\theta}) = \frac{1}{2}\ln \left[\left(R - \tilde{r}\cos\left(\tilde{\theta}\right) \right)^2 + \left(\tilde{r}\sin\left(\tilde{\theta}\right)\right)^2\right] + \phi_{1-}(R) ,
\end{equation}
where $(\tilde{r}, \tilde{\theta})$ is the polar coordinate of the microlens.

Now, let us turn our attention to the case where microlens are uniformly distributed in the lens plane with $\kappa_{*}$.
The average micro time delay at the boundary $R$ is
\begin{equation}
\begin{split}
\label{eq:Avet}
\left<\tau_\text{m} \right>(R) &= \int_0^{2\pi} \int_0^R \frac{\tau_\text{m}(R,\tilde{r},\tilde{\theta})\tilde{r}}{\pi R^2}\mathrm{d} \tilde{r} \mathrm{~d} \tilde{\theta} = 0 .
\end{split}
\end{equation}
In addition, the time delay fluctuations at the boundary $R$ induced by all the microlenses are
\begin{equation}
\begin{split}
\label{eq:flua}
\sigma_t(R) & = \sqrt{\frac{\kappa_{*}}{\pi}\int_0^{2\pi} \int_0^R\tau_\text{m}(R,\tilde{r},\tilde{\theta})^2 \tilde{r}\mathrm{d} \tilde{r} \mathrm{~d} \tilde{\theta}} = \sqrt{\frac{\kappa_*}{\pi}} R = \sqrt{\frac{\mathrm{N}_{*}}{\pi}} ,
\end{split}
\end{equation}
where $\mathrm{N}_{*}$ is the number of microlenses within the radius $R$.
One can find that it is monotonically increasing as a function of lens plane radius or the number of microlenses.

In order to verify the correctness of Eq.~(\ref{eq:flua}), here, we simulate the microlens field $t_\text{m}(\boldsymbol{x},\boldsymbol{x}^{i})$ for four different configurations.
The parameter values for simulation are listed in Table~\ref{ta:FluaTest}.
We randomly realize fifteen sky maps for each configuration, and then calculate the average standard deviation $\sigma_\text{num}$ of these fifteen sky maps.
The relevant results are also listed in Table~\ref{ta:FluaTest}.
Fig.~\ref{fig:map} shows three sky maps using the ``test $2$'' configuration.
The color bar represents the time delay values in the unit of seconds.

According to Table~\ref{ta:FluaTest}, one can find that the perturbation of the microlens calculated using our theoretical formula Eq.~(\ref{eq:flua}) is in good agreement with the numerical standard deviation $\sigma_\text{num}$ and just has a $\sim20\%$ underestimation.
These results demonstrate that Eq.~(\ref{eq:flua}) is reliable and can be used to estimate the time delay perturbation caused by microlensing.
It should be noted here that in order to accelerate the calculation speed of $t_\text{m}(\boldsymbol{x},\boldsymbol{x}^{i})$ further, we use the Taylor expansion method to approximate the far-field microlenses potential, see Wambsganss (1990)~\cite{Wambsganss1990}, Chen et al. (2021)~\cite{2021xuechunchen} and Zheng et al. (2022)~\cite{zheng2022} for more details.

\begin{table}
  \centering
  \caption{\label{ta:FluaTest} This table listed the microlens field parameter values used in simulations.
  $R$ is the lens plane boundary.
  $\kappa_{*}$ is the microlens convergence.
  $\mathrm{N}_{*}$ is the microlenses number within the lens plane boundary.
  $\sigma_\text{num}$ is the standard deviation of the time delay calculated using the histogram statistic method.
  $\sigma_t(R)$ is the microlenses fluctuation calculated using Eq.~(\ref{eq:flua}).
  Here, we set the mass of microlens as $1~\mathrm{M}_\odot$.}
   \begin{tabular}{|c|ccccc} 
    \hline
     & $R $ & $\kappa_{*}$ & $\mathrm{N}_{*}$ & $\sigma_\text{num}$ & $\sigma_t(R)$ \\ 
    \hline
    \($\text{test} 1$\) & $500$ & $0.06$ & $19099$ & $0.0025$  & $0.002$\\
    \($\text{test} 2$\) & $500$ & $0.04$ & $12732$ & $0.002$  & $0.0016$\\
    \($\text{test} 3$\) & $250$ & $0.04$ & $3183$ & $0.001$  & $0.0008$\\
    \($\text{test} 4$\) & $250$ & $0.16$ & $12732$ & $0.002$  & $0.0016$\\
    \hline
  \end{tabular}
\end{table}

Back to the point, and then we define a signal-to-noise ratio $\mathrm{SNR}(R)$, which indicates the ratio of the macro time delay to the time delay disturbance caused by the microlens at the radius $R$, to estimate the required lens plane boundary. 
\begin{equation}
\label{eq:Precision}
\mathrm{SNR}(R) =  \frac{\min{|\tau_\text{s}(\kappa, \gamma, \boldsymbol{R})}|}{\sigma_t(R)} = \frac{\min{(|1-\kappa-\gamma|, |1-\kappa+\gamma|)} R}{2 \times \sqrt{\frac{\kappa_*}{\pi}}} \\,
\end{equation}
where we use $\min{(|\tau_\text{s}(\kappa, \gamma, R, \theta=0)|,|\tau_\text{s}(\kappa, \gamma, R, \theta=\frac{\pi}{2})|)}$ to approximate $\min{|\tau_\text{s}(\kappa, \gamma, \boldsymbol{R})|}$ in the second equation.

One can find that with the increase of $R$, the proportion of microlens disturbance in the macro time delay decreases gradually.
This behaviour also means that with the increase of $R$, $\left|\tilde{F}_\text{residual}\left(t\right)\right| = \left|\tilde{F}\left(t\right) -  \tilde{F}_\text{s}\left(t\right)\right|$ will gradually approach $0$.
Here we define two new variables, $\tilde{F}_\text{residual}\left(t\right)$, which characterizes the microlensing fluctuation in the time domain magnification factor, and $\tilde{F}_\text{s}\left(t\right)$, which stand for the time domain magnification factor caused by smooth gravitational potential and geometry.
According to the above discussion, we can determine the boundary by using a minimum signal-to-noise ratio $\mathrm{SNR}_\mathrm{min}$, that is,
\begin{equation}
\begin{split}
\label{eq:minimumRT}
&R_\text{min} = \frac{2 \times \sqrt{\frac{\kappa_*}{\pi}} \times \mathrm{SNR}_\mathrm{min}}{\min{(|1-\kappa-\gamma|, |1-\kappa+\gamma|)}} \\
&|t_\text{min}| = \frac{k}{2}\min{(|1-\kappa-\gamma|, |1-\kappa+\gamma|)} R_\text{min}^2 ,
\end{split}
\end{equation}
where $R_\text{min}$ and $|t_\text{min}|$ are minimum lens plane boundary and time delay.
It is worth mentioning that the boundary $R_\text{min}$ is different from the contour $\boldsymbol{x_c}$, which may not be circular. 
Here, we use a circular lens plane boundary $R_\text{min}$ for simplicity.

However, $\boldsymbol{x_c}$ can be used as a statistical boundary where we count the pixels.
Since counting in this boundary can reduce the influence of boundary imperfection on $\tilde{F}(t)$.
In detail, the coordinates ($x_1$, $x_2$) of this statistical boundary need to satisfy
\begin{equation}
\label{eq:boundary}
t_\text{min} = \frac{k}{2}\left((1-\kappa+\gamma)x_1^2+(1-\kappa-\gamma)x_2^2\right) .
\end{equation}
The following sections will show that the statistical boundary is inside the lens plane boundary.

\subsection{Type I image}
\label{subsec:TypeI}
For Type I macro-image ($1-\kappa-\gamma > 0$, $1-\kappa+\gamma > 0$), we define $2/(k (1-\kappa+\gamma))=2 a^{2}$, $2/(k (1-\kappa-\gamma))=2 b^{2}$.
So the time-delay due to the smooth gravitational potential and geometry can be written as 
\begin{equation}
t_\text{s}=\frac{x_{1}^{2}}{2 a^{2}}+\frac{x_{2}^{2}}{2 b^{2}} \\.
\end{equation}
One can find that this is a standard elliptic equation, and the area within the $t_{\text{s}}$ contour is
\begin{equation}
S=2 \pi a b t_{\text{s}} \\.
\end{equation}
Therefore the time domain magnification factor caused by smooth gravitational potential is constant
\begin{equation}
\label{eq:TypeISmooth}
\tilde{F}_\text{s}\left(t_{\text{s}}\right)= \frac{\mathrm{d}S}{\mathrm{d}t_{\text{s}}}=\frac{2 \pi \sqrt{\mu}}{k}, \ \text{for} \ t_{\text{s}} \geq 0\\,
\end{equation}
where $\mu$ is the magnification of the Type I macro-image without considering microlenses. 
Here, we choose the minimum time delay as zero.
After inverse Fourier transforms it, one can find $F_\text{s}(\omega) = \sqrt{\mu}$, for $\omega \neq 0$.
This result also agrees with the property of the Type I macro-image without microlens.

Inspired by this phenomenon, we propose a \textbf{Component Decomposition} (CD) Method to solve the infinite time problem.
Here, we need to mention that the infinite time problem appears only at the right end of the $\tilde{F}(t)$ for the Type I macro-image.
The process is as follows.
\begin{enumerate}[Step 1:] 
	\item Use Eq.~(\ref{eq:minimumRT}) to calculate the minimum radius $R_\text{min}$ and the minimum time delay $t_\text{min}$ for one certain microlens field with convergence $\kappa_{*}$.
    \item In order to simplify the problem, we randomly put microlenses in a square area whose coordinate range is $(-R_\text{min},R_\text{min})$. The microlens number is $N = \kappa_{*}\times\text{Area}/\pi = 4 \kappa_{*} R_\text{min}^2/\pi$.
    \item Calculate $\tilde{F}(t)$ within the statistic boundary whose coordinates satisfy Eq.~(\ref{eq:boundary}). 
    Then, $\tilde{F}_\text{residual}\left(t\right)$ is calculated by subtracting Eq.~(\ref{eq:TypeISmooth}) from $\tilde{F}(t)$ within the above time delay range $t_\text{min}$.
    \item \label{Step4} Inverse Fourier transform different components.
    \begin{equation}
    \begin{split}
    \label{eq:decompose}
    F(\omega) & = C_\omega\left[{\text{IFFT}\left(\tilde{F}_\text{residual}\left(t\right)\right) + \text{IFFT}\left(\tilde{F}_\text{s}\left(t\right)\right) } \right]
    \\ & =C_\omega \text{IFFT} \left(\tilde{F}_\text{residual}\left(t\right)\right) + \sqrt{\mu} , \quad (\omega \neq 0) .
    \end{split}
    \end{equation}
\end{enumerate}

This method has two obvious advantages.
First, the lens plane boundary introduced in section~\ref{subsec:ParaSet} guarantees $\tilde{F}_\text{residual}\left(t\right) \simeq 0$ for $t > t_\text{min}$.
Therefore, the contribution from $t > t_\text{min}$ is almost zero for $\text{IFFT} \left(\tilde{F}_\text{residual}\left(t\right)\right)$.
This property frees our result from spectral leakage caused by the convolution with the window function.
Second, this method does not add extra calculation time and has greater precision, thanks to $\text{IFFT}\left(\tilde{F}_\text{s}\left(t\right)\right) = \frac{\pi \sqrt{\mu}}{k}\text{IFFT}\left(\text{Sgn}(t) + 1\right)= \frac{\sqrt{\mu}}{C_\omega}\ (\omega \neq 0)$ being an analytic expression and contain an infinite time range.

\subsection{Type II image}
\label{subsec:TypeII}
For Type II macro-image, where eigenvalues of the Jacobian matrix have opposite signs, we define $2/(k (1-\kappa+\gamma))=2 a^{2} > 0$ and $2/(k (\kappa+\gamma - 1))=2 b^{2} > 0$, without loss of generality.
So the time-delay from smooth gravitational potential can be written as 
\begin{equation}
t_\text{s}=\frac{x_{1}^{2}}{2 a^{2}}-\frac{x_{2}^{2}}{2 b^{2}} .
\end{equation}
Here, we choose the zero point of time delay at the lens plane origin.

\begin{figure}
	\centering 
	\includegraphics[width=0.4\textwidth]{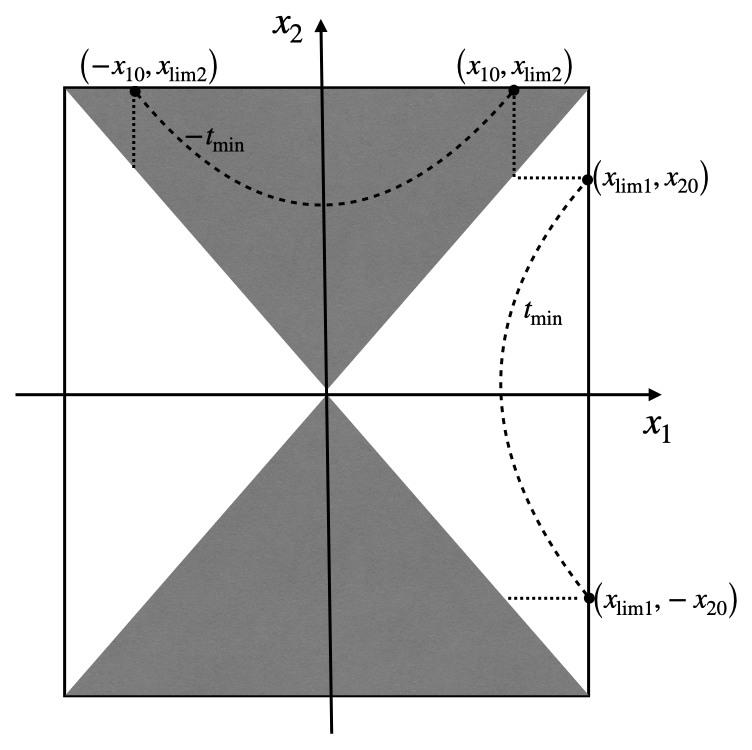}
	\caption{This figure is an illustration of Type II macro image plane. 
	The gray area represents the first part, and the white area represents the second part, as introduced in section~\ref{subsec:TypeII}. 
	Dashed curves are the contours corresponding to the time delay $|t_\text{min}|$. 
	Dotted together with Dashed curves make up the statistical boundary in different parts.
	$x_{\text{lim}1}$, $x_{\text{lim}2}$ are the lens plane boundary coordinates as discussed in section~\ref{subsec:TypeII}.
	}
	\label{fig:sketch}
\end{figure}

The asymptotes can divide the lens plane into four regions, in which the time delays of the upper and lower parts (hereafter referred to as the first part) are less than $0$, and the time delays of the left and right parts (hereafter referred to as the second part) are greater than $0$.
Fig.~\ref{fig:sketch} shows the schematic diagram of the Type II macro image plane, and the quantities therein will be introduced in the subsequent text.

In the first part, the area between the contour and the asymptote is
\begin{equation}
S = \lim_{x_{10}\to+\infty}2\int_{-x_{10}}^{x_{10}} \sqrt{\frac{b^2}{a^2}x_1^2 - 2b^2t_\text{s}} - \frac{b}{a}x_1 ~\mathrm{d}x_1 .
\end{equation}
The time domain magnification factor is 
\begin{equation}
\begin{split}
\label{eq:TypeIITimeMag}
\tilde{F}_\text{s}\left(t_\text{s}\right) &= \lim_{x_{10}\to+\infty} -2 \frac{\sqrt{\mu}}{k}\left[\ln 2+2 \ln a+\ln |t_\text{s}|-2 \ln \left(x_{10}+\sqrt{2 a^{2}|t_\text{s}|+x_{10}^{2}}\right)\right] \\
 &= -2 \frac{\sqrt{\mu}}{k}\left[\ln |t_\text{s}|\right] + \text{C}
 \end{split}
\end{equation}
where C is an infinite constant.

Similarly, the area between the contour and the asymptote in the second part is
\begin{equation}
S = \lim_{x_{20}\to +\infty}2\int_{-x_{20}}^{x_{20}} \sqrt{\frac{a^2}{b^2}x_2^2 + 2a^2t_\text{s}} - \frac{a}{b}x_2 ~\mathrm{d} x_2 .
\end{equation}
And the time domain magnification factor is 
\begin{equation}
\begin{split}
\label{eq:TypeIITimeMag2}
\tilde{F}_\text{s}\left(t_\text{s}\right) &=\lim_{x_{20}\to+\infty}-2 \frac{\sqrt{\mu}}{k}\left[\ln 2+2 \ln b+\ln t_\text{s}-2 \ln \left(x_{20}+\sqrt{2 b^{2} t_\text{s}+x_{20}^{2}}\right)\right] \\
&=-2 \frac{\sqrt{\mu}}{k}\left[\ln t_\text{s}\right] + \text{C}
\end{split}
\end{equation}
Therefore, in the whole lens plane, the time domain magnification factor can be written as
\begin{equation}
\tilde{F}_\text{s}\left(t_\text{s}\right) = -2\frac{\sqrt{\mu}}{k}\left[\ln|t_\text{s}|\right] + \text{C} .
\end{equation}

Unlike Type I, one can find that the boundaries Eq.~(\ref{eq:boundary}) are divergent hyperbola curves.
Therefore, we need to delimit a finite region so that the result $\tilde{F}_\text{s}\left(t\right)$ obtained in this region is only one constant $C'$ different from that obtained in the infinite region.

Through Eq.~(\ref{eq:minimumRT}), we can determine the minimum time delay $|t_\text{min}|$.
In the first part, in order to make Eq.~(\ref{eq:TypeIITimeMag}) hold in a finite lens plane area, the following condition needs to be satisfied.
\begin{equation}
\sqrt{\frac{b^2}{a^2}x_{10}^2 + 2b^2|t_\text{min}|} - \frac{b}{a}x_{10} < \epsilon_1^\prime .
\end{equation}
This inequality means that at $x_{10}$, the ordinate difference between the hyperbola curve for time delay $|t_\text{min}|$ and the asymptote is less than $\epsilon_1^\prime$.
By solving this inequality, we can get
\begin{equation}
\label{eq:x10}
x_{10} > -\frac{a\epsilon_1}{2b} + \frac{ab|t_\text{min}|}{\epsilon_1} .
\end{equation}
Considering this condition, the last term in Eq.~(\ref{eq:TypeIITimeMag}) can be written as 
\begin{equation}
C_\text{finite1} = -\frac{2\sqrt{\mu}}{k}[\ln2 + 2\ln a-2\ln(2x_{10}+\frac{a}{b}\epsilon_1^\prime)] ,
\end{equation}
which is a time independent constant.

Similarly, in the second part, the following condition needs to be satisfied
\begin{equation}
\sqrt{\frac{a^2}{b^2}x_{20}^2 + 2a^2|t_\text{min}|} - \frac{a}{b}x_{20} < \epsilon_2^\prime .
\end{equation}
Therefore
\begin{equation}
\label{eq:x20}
x_{20} > -\frac{b\epsilon_2}{2a} + \frac{ab|t_\text{min}|}{\epsilon_2} .
\end{equation}
In this condition, the last term in Eq.~(\ref{eq:TypeIITimeMag2}) can be written as
\begin{equation}
C_\text{finite2} = -\frac{2\sqrt{\mu}}{k}\left[\ln2 + 2\ln b-2\ln(2x_{20}+\frac{b}{a}\epsilon_2^\prime)\right] .
\end{equation}
It is worth noting that to analogy the results of the infinite lens plane as much as possible. 
We need to make $C_\text{infinite1} = C_\text{infinite2}$.
One can find that the above condition can be satisfied only by making $\epsilon_2^\prime = \frac{a}{b}\epsilon_1^\prime$.
Finally, one can get the lens plane boundary coordinate $x_{\text{lim}1}$ and $x_{\text{lim}2}$ by inserting $x_{20}$ and $x_{10}$ into their corresponding hyperbola curves.
In addition, one can also find that the diagonals of this lens plane are precisely the asymptotic curves of the hyperbola.
These coordinate relationships are shown in Fig.~\ref{fig:sketch}.

Now we will introduce the CD method for Type II.
Here, we need to mention that, unlike Type I, the infinite time problem of Type II appears at both ends of the curve.
\begin{enumerate}[Step 1:] 
    \item Randomly place microlenses ($N = \kappa_{*}\times\text{Area}/\pi = 4\kappa_{*} x_{\text{lim}1}x_{\text{lim}2}/\pi$) in a rectangle lens plane whose abscissa and ordinate ranges are $(-x_{\text{lim}1},x_{\text{lim}1})$ and $(-x_{\text{lim}2},x_{\text{lim}2})$.
    \item Calculate $\tilde{F}(t)$ within the hyperbola curves whose boundary coordinates satisfy Eq.~(\ref{eq:boundary}) and $x_1\in[-x_{10},x_{10}]$ for $t_\text{s}<0$, $x_2\in[-x_{20},x_{20}]$ for $t_\text{s}>0$. 
    Then, use the first equations in Eq.~(\ref{eq:TypeIITimeMag}) and Eq.~(\ref{eq:TypeIITimeMag2}) to calculate $\tilde{F}^\prime_\text{s}\left(t\right)$ in the first part and the second part, respectively.
    \item Calculate $\tilde{F}_\text{residual}\left(t\right)$ using $\tilde{F}_\text{residual}\left(t\right) = \tilde{F}\left(t\right) - \tilde{F}^\prime_\text{s}\left(t\right)$.
    \item Inverse Fourier transform different components.
    \begin{equation}
    \begin{split}
    \label{eq:decomposeII}
    F(\omega) & = C_\omega\left[{\text{IFFT}\left(\tilde{F}_\text{residual}\left(t\right)\right) + \text{IFFT}\left(\tilde{F}_\text{s}\left(t\right)\right) } \right]
    \\ & =C_\omega \text{IFFT}\left(\tilde{F}_\text{residual}\left(t\right)\right) - i\times\text{Sgn}(\omega) \sqrt{\mu} , \quad (\omega \neq 0).
    \end{split}
    \end{equation}
\end{enumerate}

\subsection{Type III image}
\label{subsec:TypeIII}
Type III is almost the same as Type I, except that the infinite time problem appears at the left end of the curve because eigenvalues of the Jacobian matrix are less than $0$.
we define $-2/(k (1-\kappa+\gamma)))=2 a^{2}$, $-2/(k (1-\kappa-\gamma))=2 b^{2}$.
Therefore the time delay is
\begin{equation}
t_\text{s}= - \left[\frac{x_{1}^{2}}{2 a^{2}}+\frac{x_{2}^{2}}{2 b^{2}}\right] \\,
\end{equation}
and we choose the maximum time delay as zero.
The area within the $t_\text{s}$ contour is
\begin{equation}
S=2 \pi a b|t_\text{s}| \\.
\end{equation}
Therefore, the time domain magnification factor is the same as that of Type I but has a different domain of definition
\begin{equation}
\tilde{F}_\text{s}\left(t_\text{s}\right)=2 \pi a b=\frac{2 \pi \sqrt{\mu}}{k}, \ \text{for} \ t_\text{s} \leq 0\\.
\end{equation}
Moreover, Type III uses the same CD method as type I, the only difference is Eq.~(\ref{eq:decompose})
\begin{equation}
F(\omega)=C_\omega \text{IFFT} \left(\tilde{F}_\text{residual}\left(t\right)\right) - \sqrt{\mu} , \quad (\omega \neq 0) \\.
\end{equation}

\section{Result for a point-mass}
\label{sec:ResOnePoint}
This section demonstrates the methodology by using a point mass model embedded in Type I, Type II, and Type III macro images. 
We set the lens mass $\mathrm{M_{micro}} = 100~\mathrm{M}_\odot$.
The lens redshift $z_L = 0.5$ and source redshift $z_s = 1$.
We will use the same lens and source redshift in the following text unless otherwise specified.

In order to test the accuracy of our algorithm, we need to compare it with the ``standard result''.
Fortunately, we find that for a point mass lens embedded in external convergence but without external shear, Eq.~(\ref{eq:DiffInter}) can be integrated analytically.
Hereinafter, we give this analytical expression for the first time (the calculation details are similar to Deguchi \& Watson (1986)~\cite{Shuji1986}).
\begin{equation}
\label{eq:embededin}
\begin{split} 
F(w,\boldsymbol{y}=0,\boldsymbol{x})=&-\frac{1}{\lambda^2 w} 2^{-2-\frac{i w}{2}}|w|^{-1+\frac{i w}{2}} \Gamma\left(1-\frac{i w}{2}\right)|\lambda|^{-1+\frac{i w}{2}} e^{\frac{i w}{2}\left(\phi_{m} + \lambda(x_1^2 + x_2^2)\right)} \quad\left((w+2 i)\left|\lambda\right|^{3}|w|^{3}\left(x_{1}^{2}+x_{2}^{2}\right)\right.
\\ &\left(\sinh \left(\frac{\pi w}{4}\right) \operatorname{sgn}(\lambda w)+\cosh \left(\frac{\pi w}{4}\right)\right){ }_{2} F_{3}\left(1-\frac{i w}{4}, \frac{3}{2}-\frac{i w}{4} ; 1, \frac{3}{2}, \frac{3}{2} ;-\frac{1}{16}\lambda^2 w^{2}\left(x_{1}^{2}+x_{2}^{2}\right)^{2}\right) 
\\ &\left.-4 \lambda^2w^2\left(\cosh \left(\frac{\pi w}{4}\right) \operatorname{sgn}(\lambda w)+\sinh \left(\frac{\pi w}{4}\right)\right){ }_{2} F_{3}\left(\frac{1}{2}-\frac{i w}{4}, 1-\frac{i w}{4} ; \frac{1}{2}, \frac{1}{2}, 1 ;-\frac{1}{16}\lambda^2 w^{2}\left(x_{1}^{2}+x_{2}^{2}\right)^{2}\right)\right) 
\end{split}
\end{equation}
where $w = k\omega$ is dimensionless frequency, $\mu$ is the macro magnification, $\phi_{m}$ is used to calibrate the time delay zero point, $\Gamma$ and ${}_{2}F_{3}$ are gamma function and generalized hypergeometric function, respectively.
One can find that this formula can be used to calculate the cases of Type I and Type III when $\gamma = 0$.

In the following texts, we compare our CD method with the analytical result (Eq.~(\ref{eq:embededin})) and the results of the traditional apodization method with long enough or short time delay, respectively. 

\begin{figure*}
	\centering 
	\includegraphics[width=\textwidth]{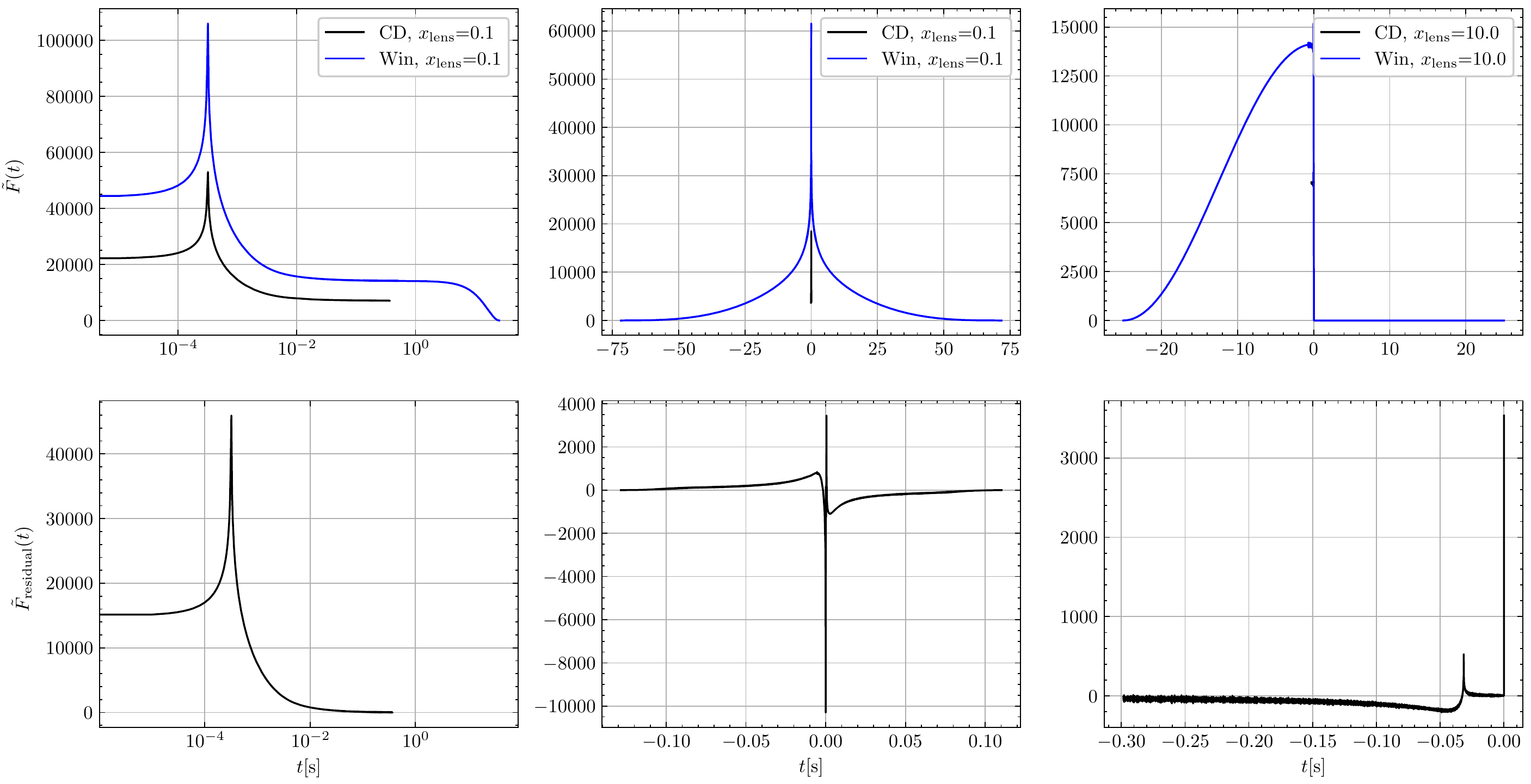}
	\caption{A point mass lens embedded in Type I ($\kappa=0.7$, $\gamma=0$), Type II ($\kappa=0.875$, $\gamma=0.325$), and Type III ($\kappa=1.3$, $\gamma=0$) macro image environment.
	The first row shows the time domain magnification factor (Eq.~(\ref{eq:TimeDomainMag})), and the second row refers to the residual time domain magnification $\tilde{F}_\text{residual}(t)$ as a function of time delay $t$.
	The first, second, and third columns stand for the result for Type I, Type II, and Type III circumstances, respectively. 
	The black curves stand for the time domain magnification factors and $\tilde{F}_\text{residual}(t)$ used in our CD method.
	The blue curves stand for the time domain magnification factors multiplied by $2\times$ Hanning window functions used in the traditional apodization method.}
	\label{fig:TimeDomainMagnificationFactor}
\end{figure*}

In studying a single microlens system, the boundary problem is relatively easy to deal with.
The microlens will only produce disturbances in its nearby area, so it is unnecessary and inappropriate to use the boundary estimation method introduced in section~\ref{subsec:ParaSet}.
In this scenario, since the time delay between geometric images caused by $100~\mathrm{M_\odot}$ microlens is on the order of $10^{-3}$ seconds, $\tilde{F}(t)$ with a length of $1$ second is sufficient.
This is a very conservative length, so we can make cuts in the original series, removing most of the tails where $\tilde{F}_\text{residual}(t)$ is $0$ and keeping only a tiny portion of the series where $\tilde{F}_\text{residual}(t)$ is $0$.
One can find that the time domain magnification factor used in our CD method is less than $0.3$ seconds, as shown in Fig.~\ref{fig:TimeDomainMagnificationFactor}.

However, for the traditional apodization method, we should at least ensure that results in the frequency range affected by the microlens can be accurately calculated.
It is because the spectrum leakage will affect the amplitude of the magnification factor, and the results at lower frequencies unaffected by microlens can be extrapolated by geometrical optical amplification $\sqrt{\mu}$.
Here, we use the dimensionless frequency to estimate the maximum time delay for apodization method~\citep{2003Takahashi,Gao:2021sxw}
\begin{equation}
w= k\omega \sim \frac{R_\text{S}}{\lambda} \\.
\end{equation}
$R_\text{S}$ is the Schwarzschild radius of the microlens and $\lambda$ is the wavelength of the GW.
Here we assume that the wavelength of $\sim1000$ times the Schwarzschild radius will not be affected by the microlens. 
It is a very conservative choice in order to obtain an accurate result and compare it with the result of the CD method.
In this case, the lowest frequency of GW is $\sim0.1$Hz, which needs a time delay range of ten seconds for inverse Fourier transform.

\begin{figure*}
	\centering 
	\includegraphics[width=\textwidth]{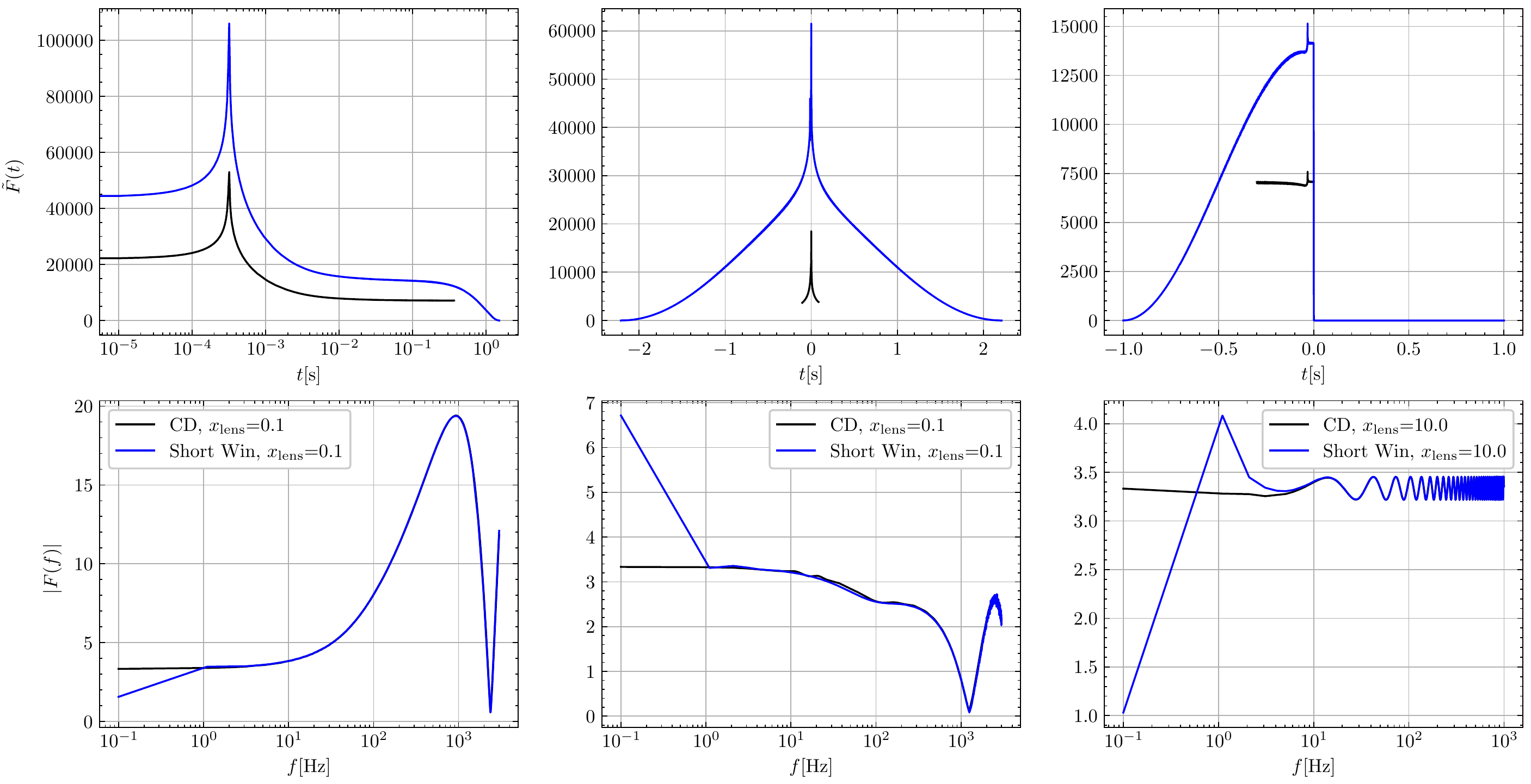}
	\caption{A point mass lens embedded in Type I ($\kappa=0.7$, $\gamma=0$), Type II ($\kappa=0.875$, $\gamma=0.325$), and Type III ($\kappa=1.3$, $\gamma=0$) macro image environment.
	The first row's black and blue curves refer to the time-domain magnification factor used in our CD and traditional apodization methods.
	The second row stands for the corresponding frequency-domain magnification factor.}
	\label{fig:ShortTime}
\end{figure*}

Fig.~\ref{fig:TimeDomainMagnificationFactor} shows time domain magnification factor results for a point mass lens embedded in Type I ($\kappa=0.7$, $\gamma=0$), Type II ($\kappa=0.875$, $\gamma=0.325$), and Type III ($\kappa=1.3$, $\gamma=0$) macro images.
The microlens coordinates are shown in the legend.
The first and second rows refer to the magnification factor and the residual magnification factor $\tilde{F}_\text{residual}(t)$ in the time domain, respectively.
The first, second, and third columns denote the results for Type I, Type II, and Type III circumstances, respectively.
We use black curves to represent the time domain magnification and $\tilde{F}_\text{residual}(t)$ used in our CD method.
Furthermore, blue curves stand for time domain magnification factors multiplied by $2\times$ Hanning window functions used in the traditional apodization method.
It is worth noting that the multiplication factor $2$ leads to different amplitudes of the black and blue curves in the first rows.
This operation is only for better identification of different curves. 
When calculating the frequency domain magnification factor using the traditional apodization method, we use the time domain magnification factor multiplied by the $1\times$ Hanning window functions.
The difference in amplitude between the black and blue curves in the first row of Fig.~\ref{fig:ShortTime} is also the same reason.

\begin{figure*}
	\centering 
	\includegraphics[width=\textwidth]{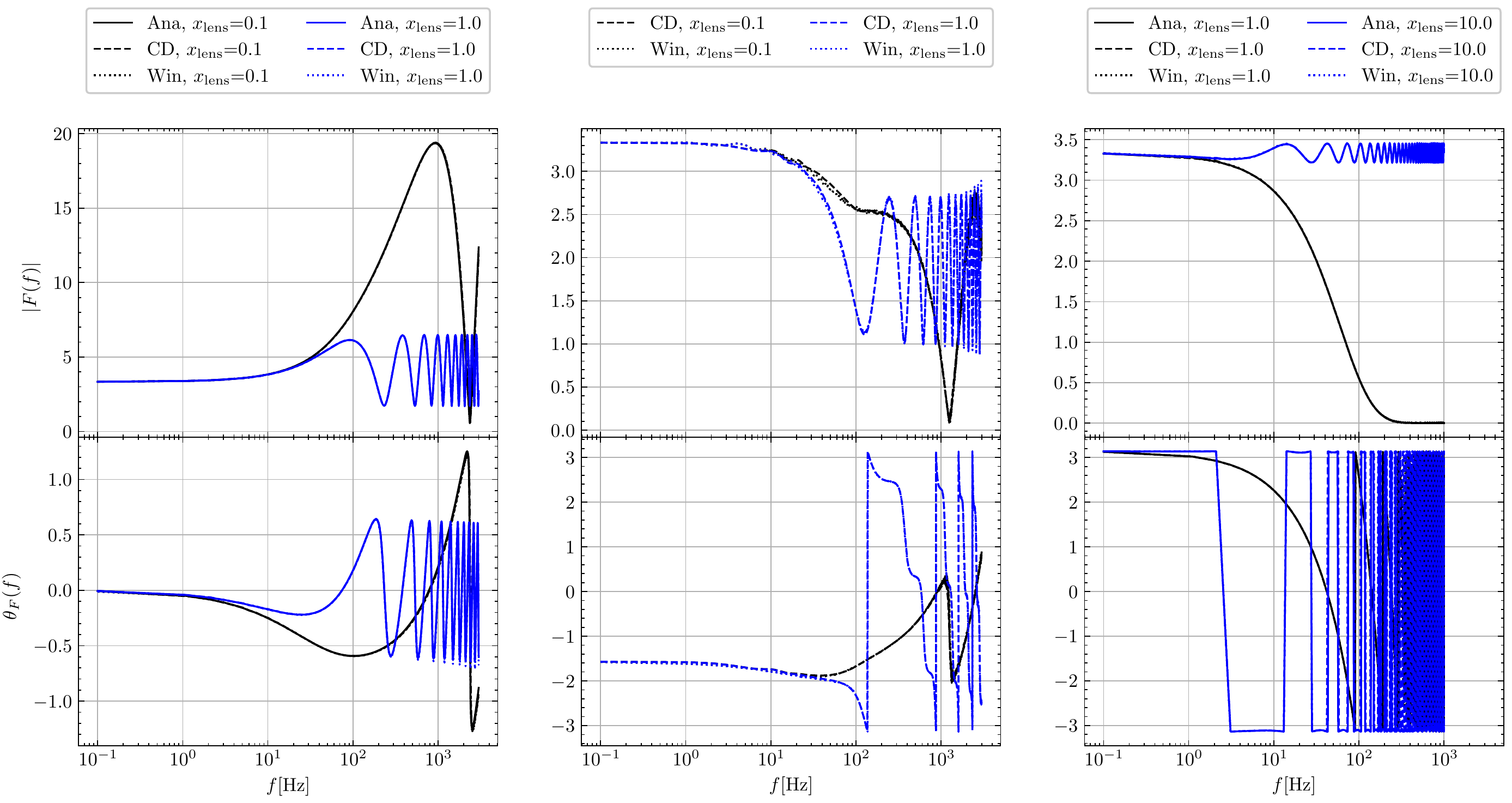}
	\caption{A point mass lens embedded in Type I ($\kappa=0.7$, $\gamma=0$), Type II ($\kappa=0.875$, $\gamma=0.325$), and Type III ($\kappa=1.3$, $\gamma=0$) macro image environment.
	The first row shows the magnification factor as a function of frequency $f$.
	The second row refers to the complex phase as a function of frequency $f$.
	The first, second, and third columns stand for the result for Type I, Type II, and Type III circumstances, respectively. 
	Solid curves in the first and third columns are analytical results using Eq.~(\ref{eq:embededin}). 
	Dashed and dotted curves represent the results using our CD and traditional apodization methods, respectively.
	Different colors stand for different lens positions, as shown in the legend.
	It is worth noting that the solid, dashed, and dotted curves of the same color are overlapped.}
	\label{fig:MagnificationFactor}
\end{figure*}

Fig.~\ref{fig:MagnificationFactor} shows the magnification factor results for the upper three macro configurations.
In the first row, we show the magnification factor as a function of frequency $f$, and the second row represents the complex phase.
The first, second, and third columns stand for the result for Type I, Type II, and Type III circumstances, respectively. 
Solid curves in the first and third columns are analytical results using Eq.~(\ref{eq:embededin}). 
Dashed and dotted curves represent the results using our CD and traditional apodization methods, respectively.
Different colors stand for different lens positions, as shown in the legend.
It is worth noting that, in the case of Type III, the coordinates of the microlens are more extensive than those in the first two cases.
Since if the microlens is too close to the Type III maximum point, there will be no micro image.
Here, we choose two different configurations to illustrate the two different situations.
One has microimages ($y=10$), and another does not ($y=1$).

One can find that the results obtained by our method are consistent with the analytical results for Type I and Type III cases with high precision.
Moreover, according to the results in the second column of Fig.~\ref{fig:MagnificationFactor}, one can also find that our method fits well with the traditional apodization method for the Type II case.
So, in summary, our method can deal with all cases of microlens embedded in macro images.

In order to further illustrate the advantages of our CD method over the traditional apodization method, we shorten the time delay length used in the traditional method to less than $10$ seconds (still longer than that used in our CD method).
Fig.~\ref{fig:ShortTime} shows the comparison results.
The black and blue curves in the first row refer to the time domain magnification factor used in our CD and traditional apodization methods, respectively. 
The second row stands for the corresponding frequency-domain results.
One can find that there are apparent errors in traditional methods at low frequencies, which is affected by frequency spectrum leakage.
Therefore, our CD method is faster and more accurate.

Finally, we find an additional result.
In the case of microlens at $y=1.0$ for Type III macro image, the magnification factor at high frequency is null since there is no micro image.
However, there are still wave effects at low frequencies.
Hence, if we receive a gravitational wave signal with a high-frequency truncation, it might be a Type III macro image affected by microlens.

\section{Result for microlens field}
\label{sec:ResMulMic}
\begin{table}
  \centering
  \caption{\label{ta:MultiPara} This table listed the lens parameter values of Type I, Type II, and Type III macro images used in simulations.
  $\kappa$ and $\gamma$ are the convergence and shear due to the macro-lens. 
  $\kappa_{*}$ is the microlens convergence.
  $\mathrm{SNR}_\mathrm{min}$ is the minimum SNR used in these calculations.
  $|t_\text{min}|$ is the minimum time delay, and $R_\text{min}$ is the corresponding lens plane boundary for Type I and Type III.
  $x_{\text{lim}1}$ and $x_{\text{lim}2}$ is the lens plane boundary for Type II.
  $x_{10}$ and $x_{20}$ are the statistic boundaries used in Eq.~(\ref{eq:TypeIITimeMag}) and Eq.~(\ref{eq:TypeIITimeMag2}).
  $\mathrm{N}_{*}$ is the microlenses number within the lens plane boundary.
  Here, we set the mass of the microlens as $1~\mathrm{M}_\odot$.}
   \begin{tabular}{|c|ccccccccccc} 
    \hline
    \($\text{Parameter}$\) & $\kappa$ & $\gamma$ & $\kappa_{*}$ & $\mathrm{SNR}_\mathrm{min}$ & $|t_\text{min}|$ & $R_\text{min}$ & $x_{\text{lim}1}$ & $x_{\text{lim}2}$ & $x_{10}$ & $x_{20}$ & $\mathrm{N}_{*}$ \\ 
    \hline
    \($\text{Type I}$\) & $0.7$  & $-0.25$ & $0.06$ & $60$ & $0.08$ & $333.3$ & $/$ & $/$ & $/$ & $/$ & $8487$ \\
    \($\text{Type II}$\) & $0.8$  & $0.25$ & $0.06$ & $60$ & $0.08$ & $/$ & $148.45$ & $445.36$ & $98.45$ & $295.36$ & $5050$ \\
    \($\text{Type III}$\) & $1.2$  & $-0.15$ & $0.06$ & $60$ & $0.08$ & $333.3$ & $/$ & $/$ & $/$ & $/$ & $8487$ \\
    \hline
  \end{tabular}
\end{table}
This section applies our CD method to a more realistic case where the microlens field is embedded in external gravitational potential.
The parameter values for calculation are listed in Table~\ref{ta:MultiPara}.
Here, we set the minimum signal-to-noise ratio $\mathrm{SNR}_\mathrm{min}$ to $60$.
For Type I and Type III, we get that the corresponding lens plane radius $R_\text{min}$ is $333.3$ in the unit of the Einstein radius, and the minimum time delay $|t_\text{min}|$ is $0.08s$, by using Eq.~(\ref{eq:minimumRT}).
Then we directly simulated square lens planes with the side length of $2R_\text{min}$, as discussed in section~\ref{subsec:TypeI} and section~\ref{subsec:TypeIII}.
However, in the case of Type II macro image, first, we calculated that the abscissa limit $x_{10}$ and ordinate limit $x_{20}$ of statistic boundaries were $98.45$ and $295.36$ using Eq.~(\ref{eq:x10}) and Eq.~(\ref{eq:x20}).
After that, one could get the corresponding rectangular lens plane abscissa limit $x_{\text{lim}1} = 148.45$ and ordinate limit $x_{\text{lim}2} = 445.36$ by inserting $x_{20}$ and $x_{10}$ into there corresponding hyperbola curves.
One can look up Fig.~\ref{fig:sketch} for a better understanding.

For the three configurations described above, we use a high enough microlens density $\kappa_{*}$ so that the critical curves of microlenses can intersect with each other.
The reason for this is to produce multiple geometric optical micro images with high magnification so that apparent interference will occur at low frequencies.
The selection criterion of $\kappa_{*}$ we use is $\kappa_{*}>\min({|1-\kappa-\gamma|, |1-\kappa+\gamma|})$, described in detail in Oguri et al. (2018)~\cite{2018PhRvD..97b3518O}.
In order to better feel the role of wave optics in the microlens field and to test the validity of our calculation, we compare the results with the geometric optics approximation.

Geometric optics approximation is an approximated behavior of wave optics at a high frequency where the Fresnel-Kirchhoff diffraction integral Eq.~(\ref{eq:DiffInter}) is a rapidly oscillating function.
Therefore the integration result mainly comes from regions near the stationary points of the lens plane surface, in other words, the image points.
The geometric optics approximation formula is~\citep{10.1143/PTPS.133.137,2003Takahashi}
\begin{equation}
\label{eq:GeoDiff}
F(\omega)=\sum_{j}\left|\mu_{j}\right|^{1 / 2} \exp \left(i \omega t_{j}-i \pi n_{j}\right) \\,
\end{equation}
where $\mu_{j}$, $t_{j}$ and $n_{j}=0,1/2,1$ is the magnification, time-delay and image type of $j$th micro image.

\begin{figure}
	\centering 
	\includegraphics[width=\columnwidth]{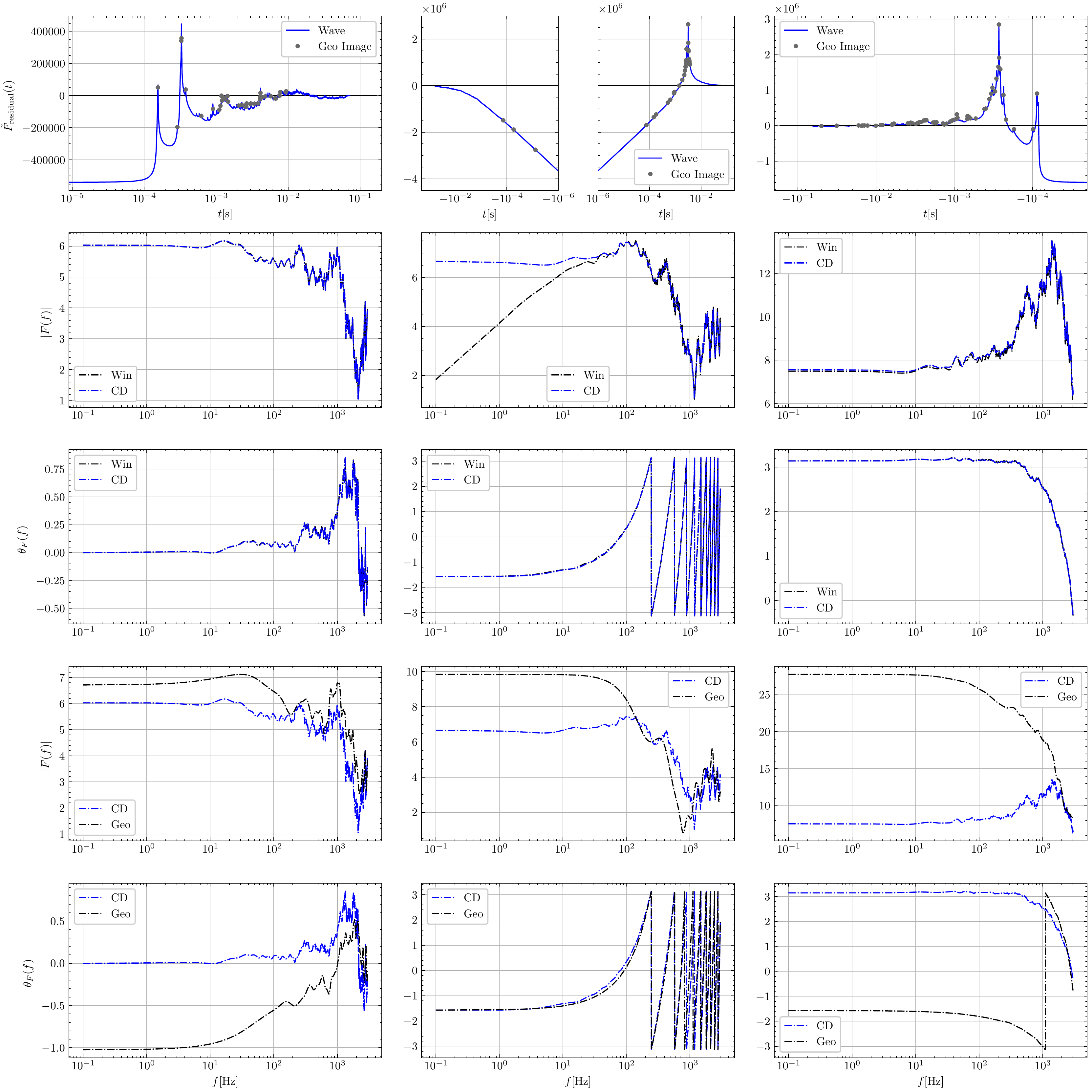}
	\caption{This figure shows the magnification results of the microlens field embedded in Type I (the leftmost column), Type II (the middle column), and Type III (the rightmost column) macro image environments, respectively.
	The parameters are listed in Table~\ref{ta:MultiPara}.
	The top panel shows the time domain magnification factor.
	The gray points represent the geometric optics microimages found by the ray-tracing method~\citep{2021xuechunchen,zheng2022}.
	The second and third rows show the comparison results of the CD method (blue dash-dotted curves) and the traditional apodization method (black dash-dotted curves).
	The forth and fifth rows show the comparison results of the CD method (blue dash-dotted curves) and geometrical approximation method (black dash-dotted curves).
	}
	\label{fig:Multi}
\end{figure}

In Fig.~\ref{fig:Multi}, we show the result for three different macro image types.
The top panel shows the time domain magnification factor.
The leftmost panel represents Type I, the middle panels represent Type II, and the rightmost panel represents Type III.
The gray points represent the geometric optics microimages found by the ray-tracing method introduced in Chen et al. (2021)~\cite{2021xuechunchen} and Zheng et al. (2022)~\cite{zheng2022}.
First, one can find that the right end for Type I, the left end for Type III and both ends for Type II gradually approach zero, proving the lens plane boundary estimation method is valid.
Second, almost every image point at the peak of the magnification factor curve. 
Therefore, our pixel counting and ray-tracing results are verified to be correct mutually.
However, some peaks have no corresponding images, lost due to the tiny image magnifications or small space separation between images.

The second and third rows show the comparison results of the CD and traditional apodization methods, and the black and blue curves in the third row overlap entirely.
One can find that the two methods agree very well in the cases of Type I and Type III.
However, for the case of type II, the traditional apodization method results have non-physical deviation at low frequency.
These phenomena are caused by the spectrum leakage of the discrete inverse Fourier transform.
In addition, it is worth mentioning that when using the apodization method, the time series is extended to $50$ seconds (for Type II, the time series is extended to $100$ seconds).
Therefore, the calculation time is $500\sim1000$ times longer than that of our CD method.
From these aspects, our CD method is faster and more accurate than the traditional method in microlens field circumstances.

The forth and fifth rows show the comparison results of the CD method and geometrical approximation method.
One can find that wave optics and geometrical optics are inconsistent at high and low frequencies. 
Inconsistencies at low frequencies have long been found~\citep[e.g.,][]{2003Takahashi,2019Diego,Cheung:2020okf,Yeung:2021roe,2021Anuj}, but they should be consistent in principle at high frequencies. 
The reason for such inconsistencies at high frequencies may be the incompleteness of geometric optical images.


\begin{figure}
	\centering 
	\includegraphics[width=\columnwidth]{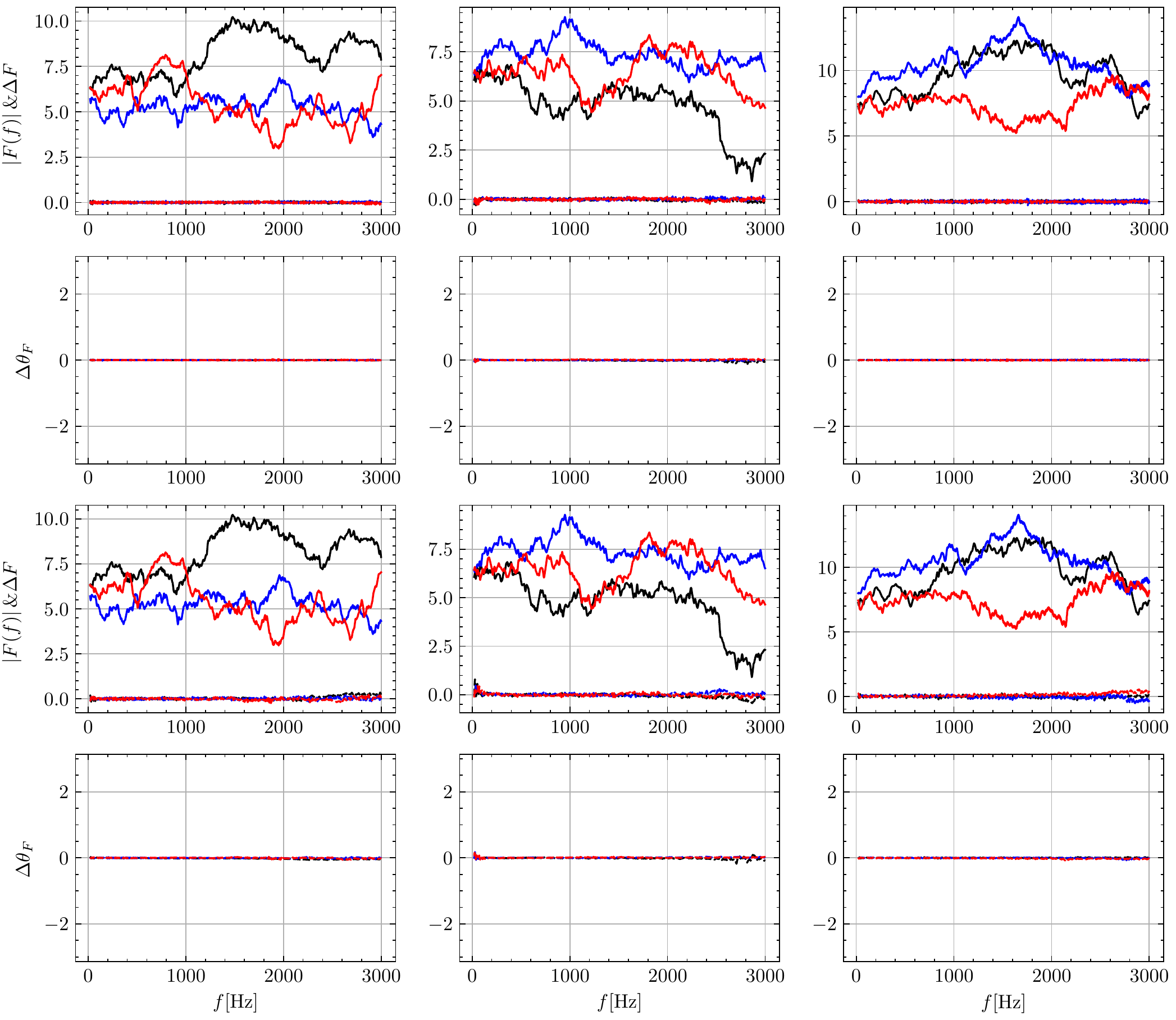}
	\caption{This figure tests the convergence of the results for $\text{SNR}_\text{min}=60$.
    The first, second, and third columns stand for the minimum, saddle, and maximum results, respectively.
    The three curves with higher amplitudes in the first row represent $|F(f)|$. 
    The parameters used are the same as those in Table~\ref{ta:MultiPara}, except for the positions of the microlenses.
    Different colors stand for different realizations.
    The three curves in the vicinity of $0$ represent the corresponding $|F(f)|$ differences between $\text{SNR}_\text{min}=60$ and $\text{SNR}_\text{min}=80$.
    The second row shows the $\theta_{F}(f)$ differences between $\text{SNR}_\text{min}=60$ and $\text{SNR}_\text{min}=80$.
    Similar to the first two rows, the third and forth rows show the differences between $\text{SNR}_\text{min}=60$ and $\text{SNR}_\text{min}=50$.
    }
	\label{fig:TypeIIShort}
\end{figure}

Moreover, we found that even for the microlens field with microlens mass as small as one solar mass, there will be noticeable interference phenomena at low frequency ($<1000$Hz).
According to the study of Takahashi \& Nakamura (2003)~\cite{2003Takahashi}, we know that the characteristics of this interference should appear around $(10^4\sim10^5)$Hz in the case of $1~\mathrm{M_{\odot}}$ isolated point lens.
This phenomenon appears in advance in the microlens field because some small microlenses close together to form a large mass of microlens.
For the phase, we found that the existence of microlens produces significant changes relative to the macro image at high frequency, which may have a significant impact on using the Morse phase information to find the macro lens image~\citep{Dai:2017huk,Dai:2020tpj} or study the higher-order modes~\citep{Wang:2021kzt,Janquart:2021nus,Vijaykumar:2022dlp}.

Finally, in Fig.~\ref{fig:TypeIIShort}, we test the stability of our algorithm by increasing and decreasing $\text{SNR}_\text{min}$.
First, we randomly realized three sky maps for each macro-image type according to the parameters ($\kappa$,$\gamma$,$\kappa_*$) in Table~\ref{ta:MultiPara} and calculated the results with $\text{SNR}_\text{min}=60$. 
Then we calculate the results of increasing $\text{SNR}_\text{min}$ to $80$ and decreasing it to $50$ and compare them with the results of $\text{SNR}_\text{min}=60$, respectively. 
We found that there is noticeable difference between $\text{SNR}_\text{min}=60$ and $\text{SNR}_\text{min}=50$ at the high frequency end. 
However, the results are relatively stable by comparing $\text{SNR}_\text{min}=60$ with $\text{SNR}_\text{min}=80$. 
Therefore, the $\text{SNR}_\text{min}=60$ is close to the minimum signal-to-noise ratio of getting a stable result.

\section{Summary and discussion}
\label{sec:SumDis}
An initial objective of this project was to provide a reliable tool and theoretical guidance for the study of the gravitational lensing wave effect of GW, especially for the microlens field scenario.
We developed a fast and accurate \textbf{Component Decomposition} (CD) method based on Ulmer \& Goodman (1995)~\cite{UG95} and Diego et al. (2019)~\cite{2019Diego} to calculate the Fresnel-Kirchhoff diffraction integral.
Compared with the traditional apodization method used in Diego et al. (2019)~\cite{2019Diego} and Mishra et al. (2021)~\cite{2021Anuj}, our method has two advantages.
First, the boundary estimation criterion introduced in section~\ref{subsec:ParaSet} guarantees $\tilde{F}_\text{residual}\left(t\right) \simeq 0$ for high $t$.
This property makes the result of discrete inverse Fourier transform of $\tilde{F}_\text{residual}(t)$ with time delay greater than $t_\text{min}$ almost zero.
Therefore, we do not have to worry about the spectrum leakage caused by truncation like the traditional apodization method and do not have the infinite time discrete Fourier transform problem.
Second, because spectral leakage can cause information contamination at low frequencies~\citep{2019Diego,2021Anuj}, the time domain magnification factor used in the traditional apodization method needs to be long enough so that information in the frequency range affected by microlenses is not contaminated.
This feature would consume more CPU time and memory.
However, our method only needs to include a relatively short time according to the boundary estimation criteria introduced in section~\ref{subsec:ParaSet}.
Overall, it is conservatively estimated that our method is $\sim500$ times faster than the traditional apodization method and has higher accuracy.

In addition, we give a quantitative estimation standard of the lens plane boundary (Eq.~(\ref{eq:minimumRT})) for studying the microlensing wave effect.
More specifically, to estimate how long time delay needs to be included to make the bias resulting from the finite time delay cut off as small as possible.

In general, we proposed a faster, more accurate, and complete algorithm equipped with a quantitative estimation standard of the lens plane boundary for studying the wave effect of microlenses embedded in different strong lensing images. 
Therefore, this method will facilitate future studies of the gravitational lensing wave effects of GWs and other long electromagnetic waves.

\vspace*{2mm} \Acknowledgements{\bahao  }
This work is supported by the NSFC (No. U1931210, 11673065, 11273061).
We acknowledge the science research grants from the China Manned Space Project with NO.CMS-CSST-2021-A11, the Sugon Advanced Computing service platform for computing support, the cosmology simulation database (CSD) in the National Basic Science Data Center (NBSDC) and its funds the NBSDC-DB-10 (No. 2020000088).
WZ is supported by the National Key R\&D Program of China Grant No. 2021YFC2203100, NSFC No. 11903030 and 11903033, the Fundamental Research Funds for the Central Universities under Grant No. WK2030000036 and WK3440000004.



\bibliographystyle{unsrt}
\bibliography{wave_effect}

\begin{thebibliography}{10}

\bibitem{2015aLIGO}
J~Aasi, B~P Abbott, R~Abbott, T~Abbott, M~R Abernathy, K~Ackley, C~Adams,
  T~Adams, P~Addesso, and et~al.
\newblock Advanced ligo.
\newblock {\em Classical and Quantum Gravity}, 32(7):074001, Mar 2015.

\bibitem{Abbott:2016blz}
B.P. Abbott et~al.
\newblock {Observation of Gravitational Waves from a Binary Black Hole Merger}.
\newblock {\em Phys. Rev. Lett.}, 116(6):061102, 2016.

\bibitem{Acernese_2014}
F~Acernese, M~Agathos, K~Agatsuma, D~Aisa, N~Allemandou, A~Allocca, J~Amarni,
  P~Astone, G~Balestri, G~Ballardin, and et~al.
\newblock Advanced virgo: a second-generation interferometric gravitational
  wave detector.
\newblock {\em Classical and Quantum Gravity}, 32(2):024001, Dec 2014.

\bibitem{KAGRA:2018plz}
T.~Akutsu et~al.
\newblock {KAGRA: 2.5 Generation Interferometric Gravitational Wave Detector}.
\newblock {\em Nature Astron.}, 3(1):35--40, 2019.

\bibitem{TheLIGOScientific:2016pea}
B.P. Abbott et~al.
\newblock {Binary Black Hole Mergers in the first Advanced LIGO Observing Run}.
\newblock {\em Phys. Rev. X}, 6(4):041015, 2016.
\newblock [Erratum: Phys.Rev.X 8, 039903 (2018)].

\bibitem{Abbott:2016nmj}
B.~P. Abbott et~al.
\newblock {GW151226: Observation of Gravitational Waves from a 22-Solar-Mass
  Binary Black Hole Coalescence}.
\newblock {\em Phys. Rev. Lett.}, 116(24):241103, 2016.

\bibitem{Abbott:2017oio}
B.P. Abbott et~al.
\newblock {GW170814: A Three-Detector Observation of Gravitational Waves from a
  Binary Black Hole Coalescence}.
\newblock {\em Phys. Rev. Lett.}, 119(14):141101, 2017.

\bibitem{TheLIGOScientific:2017qsa}
B.P. Abbott et~al.
\newblock {GW170817: Observation of Gravitational Waves from a Binary Neutron
  Star Inspiral}.
\newblock {\em Phys. Rev. Lett.}, 119(16):161101, 2017.

\bibitem{Abbott:2017gyy}
B..~P.. Abbott et~al.
\newblock {GW170608: Observation of a 19-solar-mass Binary Black Hole
  Coalescence}.
\newblock {\em Astrophys. J.}, 851(2):L35, 2017.

\bibitem{Abbott:2017vtc}
Benjamin~P. Abbott et~al.
\newblock {GW170104: Observation of a 50-Solar-Mass Binary Black Hole
  Coalescence at Redshift 0.2}.
\newblock {\em Phys. Rev. Lett.}, 118(22):221101, 2017.
\newblock [Erratum: Phys.Rev.Lett. 121, 129901 (2018)].

\bibitem{LIGOScientific:2018mvr}
B.P. Abbott et~al.
\newblock {GWTC-1: A Gravitational-Wave Transient Catalog of Compact Binary
  Mergers Observed by LIGO and Virgo during the First and Second Observing
  Runs}.
\newblock {\em Phys. Rev. X}, 9(3):031040, 2019.

\bibitem{abbott2021gwtc2}
R.~Abbott et~al.
\newblock {GWTC-2: Compact Binary Coalescences Observed by LIGO and Virgo
  During the First Half of the Third Observing Run}.
\newblock {\em Phys. Rev. X}, 11:021053, 2021.

\bibitem{LIGOScientific:2021djp}
R.~Abbott et~al.
\newblock {GWTC-3: Compact Binary Coalescences Observed by LIGO and Virgo
  During the Second Part of the Third Observing Run}.
\newblock {\em arXiv e-prints}, page arXiv:2111.03606, November 2021.

\bibitem{2003Takahashi}
Ryuichi Takahashi and Takashi Nakamura.
\newblock Wave effects in the gravitational lensing of gravitational waves from
  chirping binaries.
\newblock {\em The Astrophysical Journal}, 595(2):1039–1051, Oct 2003.

\bibitem{Christian:2018vsi}
Pierre Christian, Salvatore Vitale, and Abraham Loeb.
\newblock {Detecting Stellar Lensing of Gravitational Waves with Ground-Based
  Observatories}.
\newblock {\em Phys. Rev. D}, 98(10):103022, 2018.

\bibitem{2014ASSL..404..333P}
Michele {Punturo}, Harald {L{\"u}ck}, and Mark {Beker}.
\newblock {\em {A Third Generation Gravitational Wave Observatory: The Einstein
  Telescope}}, volume 404 of {\em Astrophysics and Space Science Library}, page
  333.
\newblock 2014.

\bibitem{Evans:2016mbw}
Benjamin~P Abbott et~al.
\newblock {Exploring the Sensitivity of Next Generation Gravitational Wave
  Detectors}.
\newblock {\em Class. Quant. Grav.}, 34(4):044001, 2017.

\bibitem{PhysRevD.90.062003}
Zhoujian Cao, Li-Fang Li, and Yan Wang.
\newblock Gravitational lensing effects on parameter estimation in
  gravitational wave detection with advanced detectors.
\newblock {\em Phys. Rev. D}, 90:062003, Sep 2014.

\bibitem{Lai:2018rto}
Kwun-Hang Lai, Otto~A. Hannuksela, Antonio Herrera-Mart\'\i{}n, Jose~M. Diego,
  Tom Broadhurst, and Tjonnie G.~F. Li.
\newblock {Discovering intermediate-mass black hole lenses through
  gravitational wave lensing}.
\newblock {\em Phys. Rev. D}, 98(8):083005, 2018.

\bibitem{PhysRevLett.122.041103}
Sunghoon Jung and Chang~Sub Shin.
\newblock Gravitational-wave fringes at ligo: Detecting compact dark matter by
  gravitational lensing.
\newblock {\em Phys. Rev. Lett.}, 122:041103, Jan 2019.

\bibitem{2019Hannuksela}
O.~A. Hannuksela, K.~Haris, K.~K.~Y. Ng, S.~Kumar, A.~K. Mehta, D.~Keitel,
  T.~G.~F. Li, and P.~Ajith.
\newblock Search for gravitational lensing signatures in ligo-virgo binary
  black hole events.
\newblock {\em The Astrophysical Journal}, 874(1):L2, Mar 2019.

\bibitem{LIGOScientific:2021izm}
R.~Abbott et~al.
\newblock {Search for Lensing Signatures in the Gravitational-Wave Observations
  from the First Half of LIGO\textendash{}Virgo\textquoteright{}s Third
  Observing Run}.
\newblock {\em Astrophys. J.}, 923(1):14, 2021.

\bibitem{Cheung:2020okf}
Mark H.~Y. Cheung, Joseph Gais, Otto~A. Hannuksela, and Tjonnie G.~F. Li.
\newblock {Stellar-mass microlensing of gravitational waves}.
\newblock {\em Mon. Not. Roy. Astron. Soc.}, 503(3):3326--3336, 2021.

\bibitem{Yeung:2021roe}
Simon M.~C. {Yeung}, Mark H.~Y. {Cheung}, Joseph A.~J. {Gais}, Otto~A.
  {Hannuksela}, and Tjonnie G.~F. {Li}.
\newblock {Microlensing of type II gravitational-wave macroimages}.
\newblock {\em arXiv e-prints}, page arXiv:2112.07635, December 2021.

\bibitem{Meena:2019ate}
Ashish~Kumar Meena and J.~S. Bagla.
\newblock {Gravitational lensing of gravitational waves: wave nature and
  prospects for detection}.
\newblock {\em Mon. Not. Roy. Astron. Soc.}, 492(1):1127--1134, 2020.

\bibitem{2019Diego}
J.~M. Diego, O.~A. Hannuksela, P.~L. Kelly, G.~Pagano, T.~Broadhurst, K.~Kim,
  T.~G.~F. Li, and G.~F. Smoot.
\newblock Observational signatures of microlensing in gravitational waves at
  ligo/virgo frequencies.
\newblock {\em Astronomy \& Astrophysics}, 627:A130, Jul 2019.

\bibitem{2021Anuj}
Anuj Mishra, Ashish~Kumar Meena, Anupreeta More, Sukanta Bose, and
  Jasjeet~Singh Bagla.
\newblock Gravitational lensing of gravitational waves: effect of microlens
  population in lensing galaxies.
\newblock {\em Monthly Notices of the Royal Astronomical Society},
  508(4):4869–4886, Oct 2021.

\bibitem{press1992numerical}
William~H Press, William~T Vetterling, Saul~A Teukolsky, and Brian~P Flannery.
\newblock {\em Numerical Recipes Example Book (FORTRAN)}.
\newblock Cambridge University Press Cambridge, 1992.

\bibitem{levin1982procedures}
David Levin.
\newblock Procedures for computing one-and two-dimensional integrals of
  functions with rapid irregular oscillations.
\newblock {\em Mathematics of Computation}, 38(158):531--538, 1982.

\bibitem{filon1930iii}
Louis Napoleon~George Filon.
\newblock Iii.—on a quadrature formula for trigonometric integrals.
\newblock {\em Proceedings of the Royal Society of Edinburgh}, 49:38--47, 1930.

\bibitem{xiang2007efficient}
S.~Xiang.
\newblock Efficient filon-type methods for $\int_a^bf(x)\,{\rm e}^{{\rm
  i}\omega g(x)}\,{\rm d}x$.
\newblock {\em Numerische Mathematik}, (4):633--658, 2007.

\bibitem{iserles2006computation}
Arieh Iserles and Syvert~P N{\o}rsett.
\newblock On the computation of highly oscillatory multivariate integrals with
  stationary points.
\newblock {\em BIT Numerical Mathematics}, 46(3):549--566, 2006.

\bibitem{guo2020convergence}
Xiao Guo and Youjun Lu.
\newblock Convergence and efficiency of different methods to compute the
  diffraction integral for gravitational lensing of gravitational waves.
\newblock {\em Physical Review D}, 102(12):124076, 2020.

\bibitem{UG95}
Andrew Ulmer and Jeremy Goodman.
\newblock Femtolensing: Beyond the semiclassical approximation.
\newblock {\em The Astrophysical Journal}, 442:67, Mar 1995.

\bibitem{1992grlebookS}
Peter {Schneider}, J{\"u}rgen {Ehlers}, and Emilio~E. {Falco}.
\newblock {\em Gravitational Lenses}.
\newblock Springer New York, NY, 1992.

\bibitem{10.1143/PTPS.133.137}
Takahiro~T. Nakamura and Shuji Deguchi.
\newblock {Wave Optics in Gravitational Lensing}.
\newblock {\em Progress of Theoretical Physics Supplement}, 133:137--153, 01
  1999.

\bibitem{Wambsganss1990}
J.~{Wambsganss}.
\newblock PhD thesis, January 1990.

\bibitem{2021xuechunchen}
Xuechun Chen, Yiping Shu, Guoliang Li, and Wenwen Zheng.
\newblock Frbs lensed by point masses. ii. the multipeaked frbs from the point
  view of microlensing.
\newblock {\em The Astrophysical Journal}, 923(1):117, Dec 2021.

\bibitem{zheng2022}
Wenwen Zheng, Xuechun Chen, Guoliang Li, and Hou-zun Chen.
\newblock {An Improved GPU-based Ray-shooting Code for Gravitational
  Microlensing}.
\newblock {\em Astrophys. J.}, 931(2):114, 2022.

\bibitem{kayser1986astrophysical}
R~Kayser, S~Refsdal, and R~Stabell.
\newblock Astrophysical applications of gravitational micro-lensing.
\newblock {\em Astronomy and Astrophysics}, 166:36--52, 1986.

\bibitem{Shuji1986}
Shuji Deguchi and William~D. Watson.
\newblock Wave effects in gravitational lensing of electromagnetic radiation.
\newblock {\em Phys. Rev. D}, 34:1708--1718, Sep 1986.

\bibitem{Gao:2021sxw}
Zucheng {Gao}, Xian {Chen}, Yi-Ming {Hu}, Jian-Dong {Zhang}, and Shun-Jia
  {Huang}.
\newblock {A higher probability of detecting lensed supermassive black hole
  binaries by LISA}.
\newblock {\em Mon. Not. Roy. Astron. Soc.}, 512(1):1--10, May 2022.

\bibitem{2018PhRvD..97b3518O}
Masamune {Oguri}, Jose~M. {Diego}, Nick {Kaiser}, Patrick~L. {Kelly}, and Tom
  {Broadhurst}.
\newblock {Understanding caustic crossings in giant arcs: Characteristic
  scales, event rates, and constraints on compact dark matter}.
\newblock {\em Phys. Rev. D}, 97(2):023518, January 2018.

\bibitem{Dai:2017huk}
Liang {Dai} and Tejaswi {Venumadhav}.
\newblock {On the waveforms of gravitationally lensed gravitational waves}.
\newblock {\em arXiv e-prints}, page arXiv:1702.04724, February 2017.

\bibitem{Dai:2020tpj}
Liang {Dai}, Barak {Zackay}, Tejaswi {Venumadhav}, Javier {Roulet}, and Matias
  {Zaldarriaga}.
\newblock {Search for Lensed Gravitational Waves Including Morse Phase
  Information: An Intriguing Candidate in O2}.
\newblock {\em arXiv e-prints}, page arXiv:2007.12709, July 2020.

\bibitem{Wang:2021kzt}
Yijun Wang, Rico K.~L. Lo, Alvin K.~Y. Li, and Yanbei Chen.
\newblock {Identifying Type II Strongly Lensed Gravitational-Wave Images in
  Third-Generation Gravitational-Wave Detectors}.
\newblock {\em Phys. Rev. D}, 103(10):104055, 2021.

\bibitem{Janquart:2021nus}
Justin Janquart, Eungwang Seo, Otto~A. Hannuksela, Tjonnie G.~F. Li, and Chris
  Van~Den Broeck.
\newblock {On the Identification of Individual Gravitational-wave Image Types
  of a Lensed System Using Higher-order Modes}.
\newblock {\em Astrophys. J. Lett.}, 923(1):L1, 2021.

\bibitem{Vijaykumar:2022dlp}
Aditya {Vijaykumar}, Ajit~Kumar {Mehta}, and Apratim {Ganguly}.
\newblock {Detection and parameter estimation challenges of Type-II lensed
  binary black hole signals}.
\newblock {\em arXiv e-prints}, page arXiv:2202.06334, February 2022.

\end{thebibliography}





\end{document}